\documentclass[11pt]{article}
\usepackage{graphicx}

\usepackage{graphicx}
\usepackage{dcolumn}
\usepackage{amsmath}
\usepackage{units}

\input{pubboard/babarsym}

\def\Bztoksksks {\ensuremath{\Bz \to \KS\KS\KS}}

\def\Ksto00     {\ensuremath{B^0 \to \pi^0\pi^0}}

\def\Bflav {\ensuremath{B_{\mbox{flav}}}\xspace}
\def\Btag {\ensuremath{B_{\mbox{tag}}}\xspace}

\newcommand{\Ups}{\ensuremath{\Upsilon}}

\newcommand{\mmiss}{\ensuremath{m_{miss}}}
\newcommand{\mb}{\ensuremath{m_{B}}}

\newcommand{\microns}{\ensuremath{\mu{\rm m}}}

\newcommand{\cksksks} {\ensuremath{C_f}}
\newcommand{\sksksks} {\ensuremath{S_f}}

\newcommand{\zrec}{\ensuremath{z_{\CP}}}
\newcommand{\ztag}{\ensuremath{z_\mathrm{tag}}}

\newcommand{\BABARPubYear}    {06}

\newcommand{\BABARConfNumber} {18}
\newcommand{\SLACPubNumber} {11987}
\newcommand{\LANLNumber} {0000}

\def\figurebox#1#2#3{%
    \def\arg{#3}%
    \ifx\arg\empty
    {\hfill\vbox{\hsize#2\hrule\hbox to #2{\vrule\hfill\vbox to #1{\hsize#2\vfill}\vrule}\hrule}\hfill}%
    \else
    {\hfill\epsfbox{#3}\hfill}%
    \fi}

\long\def\inst#1{\par\nobreak\kern 4pt\nobreak
    {\it #1}\par\vskip 10pt plus 3pt minus 3pt}

\begin{document}

{\pagestyle{empty}

\begin{flushright}
\babar-CONF-\BABARPubYear/\BABARConfNumber \\
SLAC-PUB-\SLACPubNumber \\
hep-ex/\LANLNumber \\
July 2006 \\
\end{flushright}

\par\vskip 5cm

\begin{center}
\Large \bf   \boldmath Measurement  of Time-dependent \CP\ Asymmetries  in  \Bztoksksks\   Decays
\end{center}
\bigskip

\begin{center}
\large The \babar\ Collaboration\\
\mbox{ }\\
\today
\end{center}
\bigskip \bigskip

\begin{center}
  \large \bf Abstract
\end{center}
We present an updated measurement of the time-dependent \CP-violating 
asymmetry in \Bztoksksks\ decays based on  347  million $\Y4S\to\BB$
decays collected with the \babar\ detector at the PEP-II
asymmetric-energy $B$ factory at SLAC. 
We obtain the \CP\ asymmetries 
$\sksksks = -0.66 \pm 0.26 \pm 0.08$   and 
$\cksksks = -0.14 \pm 0.22 \pm 0.05 $,   where the first
uncertainties are statistical and the second systematic.

\vfill
\begin{center}

Submitted to the 33$^{\rm rd}$ International Conference on High-Energy Physics, ICHEP 06,\\
26 July---2 August 2006, Moscow, Russia.

\end{center}

\vspace{1.0cm}
\begin{center}
{\em Stanford Linear Accelerator Center, Stanford University, 
Stanford, CA 94309} \\ \vspace{0.1cm}\hrule\vspace{0.1cm}
Work supported in part by Department of Energy contract DE-AC03-76SF00515.
\end{center}

\newpage
} 

\begin{center}
\small

The \babar\ Collaboration,
\bigskip

%
{B.~Aubert,}
{R.~Barate,}
{M.~Bona,}
{D.~Boutigny,}
{F.~Couderc,}
{Y.~Karyotakis,}
{J.~P.~Lees,}
{V.~Poireau,}
{V.~Tisserand,}
{A.~Zghiche}
\inst{Laboratoire de Physique des Particules, IN2P3/CNRS et Universit\'e de Savoie,
 F-74941 Annecy-Le-Vieux, France }
% ...

\end{center}\newpage

\section{Introduction}
\label{sec:Introduction}

In the Standard Model (SM) of particle physics, the decays  \Bztoksksks~ are 
dominated by $b \to s\bar s s$ gluonic penguin amplitudes.
Let $2\beta_{eff}$ be the \CP-violating phase difference between \Bztoksksks~ decays with and without
mixing, and $\beta = {\rm arg}(-V_{cd}V_{cb}^{\ast}/V_{td}V_{tb}^{\ast})$ 
where $V_{ij}$ are the elements of the Cabibbo-Kobayashi-Maskawa (CKM)
quark mixing matrix~\cite{CKM}.
The difference $|\sin 2\beta-\sin 2\beta_{eff}|$ is expected to be nearly zero
using calculations in SM, with theoretical
uncertainties at the level of ${\cal O}(0.01)$~\cite{ref:glnq},
thanks to the factor  $|V_{ub}V_{us}/V_{tb}V_{ts}|$ which suppresses the other contributions  
in the SM.

On the other hand,  $b \to s \bar s s$  decays involve one-loop   
transitions, so contributions from heavy new particles entering such loops
can introduce new \CP-violating phases, and these may contribute to $\beta_{eff}$~\cite{grossman}.
The value of $\stwob$ has been measured at 
the \B factories in recent years with high precision~\cite{BaBarSin2beta,BelleSin2beta},
with a world average of $0.685 \pm 0.032$~\cite{unknown:2006bi}.

The Belle and \babar{} collaborations have already reported measurements of
\CP asymmetries for $\Bz \to \phi K^0_s$~\cite{Abe:2003yt,ref:stefanprl,ref:cc} (\CP-odd)
and \Bztoksksks\ (\CP-even)~\cite{belle3ks,Aubert:2005dy}.

The time-dependent \CP\ asymmetry is obtained by measuring 
the proper-time difference $\deltat\equiv t_{\CP}-t_\text{tag}$ between a 
fully reconstructed decay \Bztoksksks{}  and the decay of a partially reconstructed tagging
\B{} meson (\Btag ).
The expected asymmetry in the decay rate $f_+$ $(f_-)$ when the tagging meson is a 
\Bz{} (\Bzb ) is given as
\begin{equation}
  \label{eqn:td}
  f_{\pm}(\deltat) \; = \; \frac{e^{-|\deltat|/\tau_{\Bz}}}{4\tau_{\Bz}} \times  
  \left[ \: 1 \; \pm \;
   \: S_f \sin{( \deltamd\deltat)} \mp C_f \cos{( \deltamd\deltat)} \: \: \right] \; , \nonumber
\end{equation} 
where $\tau_{\Bz}$ is the \Bz~ lifetime and \deltamd~ is the \Bz-\Bzb~ mixing frequency.
The parameter \sksksks~ is non-zero if there is \CP violation in the interference
between decays with and without mixing, while a non-zero value for \cksksks~
would correspond to direct \CP violation. 
In the limit of one dominant decay amplitude in $b \to s \bar s s$  
transition, the SM predicts no direct \CP\  violation, and
that $\sksksks = -\eta_f\sin 2\beta$, within the theoretical hadronic uncertainties
already mentioned.
The \CP eigenvalue $\eta_f = +1$  for \CP-even \Bztoksksks~ decays.    

In this paper, we update our measurement of the time-dependent \CP-violating
asymmetries in the decay \Bztoksksks~ previously presented in~\cite{Aubert:2005dy}, 
using a larger data set and 
reconstructing the submode with one \KS~ decaying into $\pi^0\pi^0$.
The absence of charged decay tracks originating at the \Bz decay vertex requires special techniques 
to deal with its reconstruction~\cite{ref:k0spi0prl}. 
In addition, the final state has a definite \CP content~\cite{ref:gershon}, 
so that an angular analysis is not needed.

\section{The \babar\ detector and dataset}
\label{sec:babar}

The results presented here are based on $347.5\pm 3.8$ million $\Y4S\to\BB$
decays collected 
with the \babar\ detector at the
PEP-II asymmetric-energy $\epem$ collider, located at the Stanford Linear Accelerator Center. 
The \babar\ detector~\cite{ref:babar} provides charged-particle tracking through a
combination of a five-layer double-sided silicon microstrip
detector (SVT) and a 40-layer central drift chamber, both
operating in a \unit[1.5]{T} magnetic field.
Charged kaon and pion identification is achieved through measurements of particle energy-loss in
the tracking system and Cherenkov cone angle 
in a detector of internally reflected Cherenkov light.  A
segmented CsI(Tl) electromagnetic calorimeter (EMC) provides
photon detection and electron identification.  Finally, the
instrumented flux return of the magnet allows discrimination
of muons from pions.

\boldmath
\section{$B_{\CP}$ candidates selection}
\unboldmath
\label{sec:Analysis}

The \Bztoksksks\ candidate ($B_{\CP}$) is reconstructed by combining
three \KS candidates.  Two subsamples   of $B_{\CP}$  candidates are formed. One subsample contains candidates 
formed by three $\KS\to\pip\pim$ candidates  in an event ($B_{\CP(+-) }$), while  the other subsample is made
by candidates formed by two $\KS\to\pip\pim$  candidates and 
a third  $\KS$ reconstructed in the $ \ppz$ mode ($B_{\CP(00) }$).
The two subsamples have different signal to background ratios 
and therefore different analysis requirements were applied to obtain optimal selections.

For the  $B_{\CP(+-) }$ subsample  we reconstruct $\KS \to \pipi$ candidates from pairs of oppositely charged tracks.  
The two-track composites must originate from a common vertex  with a \pipi invariant mass
within 12 \mevcc (about 4$\sigma$) of the nominal \KS~ mass,  
and have a reconstructed flight
distance ($r_{dec}$) between 0.2 and 40.0\cm\ from the beam spot in the plane
transverse to the beam.
We also require that the reconstructed \KS~
has an angle between the
transverse flight direction and the transverse momentum vector of
less than 200\mrad. 
For each  $B_{\CP(+-) }$ 
candidate two nearly independent kinematic variables are computed,
the beam-energy-substituted mass 
$\mes=\sqrt{(s/2+{\bf p}_i \cdot {\bf p}_B)^2/E_i^2+p^2_B}$, 
and the energy difference $\DeltaE=E^*_B-\sqrt{s}/2$. 
Here, $(E_i,{\bf p}_i)$ is the
four-vector of the initial \epem{} system, $\sqrt{s} $
is the center-of-mass energy, ${\bf p}_B$ is the
reconstructed momentum of the \Bz{} candidate, and $E_B^*$ is its
energy calculated in the \epem{} rest frame. For signal decays,
the \mes{} distribution peaks near the \Bz{} mass with a
resolution of about \unit[$2.5$]{\mevcc}, and the \DeltaE{}
distribution peaks near zero with a resolution of about \unit{$14$}{\mev}.
We select  $B_{\CP(+-) }$  candidates within the window
\unit[$5.22<\mes<5.30$]{\gevcc} and
\unit[$-120<\DeltaE<120$]{\mev}, which includes the signal peak
and a ``sideband'' region for background characterization.

For the  $B_{\CP(00) }$ subsample  we  form $\piz\to\gamma\gamma$ candidates from pairs
of photon candidates in the EMC. Each photon is required to be
isolated from any charged tracks, to carry a minimum energy of 50\mev, and to 
have the expected lateral shower shape.  We reconstruct $\KS \to \ppz$
candidates from \piz pairs which form an invariant mass $480 <
m_{\ppz} < 520$\mevcc.   $B_{\CP(00) }$ candidates are constrained to
originate from the \epem interaction point using a geometric fit,
based on a Kalman Filter~\cite{treefitter}.  We make a requirement on
the consistency of the $\chi^2$ of the fit which retains 93\% of the
signal events and rejects about 49\% of other \B decays.  We extract
the $\KS\to \pipi$ decay length $L_{\KS}$ and the
invariant mass ($m_{\gamma\gamma}$) from this fit and require 
 $ 100 < m_{\gamma\gamma} < 141${\mevcc} and $L_{\KS}$ greater than $5$ times
its uncertainty. For $\KS\to \pipi$ candidates we require 0.15 $< r_{dec} <$ 60.0~\cm.

For each $B_{\CP(00) }$  candidate we compute two kinematic variables,  the
reconstructed mass \mb{} and the missing mass $\mmiss = \sqrt{(q_{\epem} -
\tilde{q}_B)^2}$, where $q_{\epem}$ is the four-momentum of the initial
\epem{} system and $\tilde{q}_B$ is the four-momentum of the
\Bztoksksks{} candidate after a mass constraint on the \Bz{} is applied. 
By construction, the linear correlation coefficient between
\mmiss{} and \mb{} vanishes.  
This combination of variables shows smaller correlation (0.9\% on 
reconstructed signal Monte Carlo events
and 1.7\% on the final data sample) and better
background suppression with respect to the equivalent kinematic variables
$\DeltaE$ and $\mes$  used for  $B_{\CP(+-) }$ candidates.   This is more relevant for  $B_{\CP(00) }$  candidates 
given  the asymmetric resolution  on these variables due to $\pi^0$ energy  reconstruction.
We select  $B_{\CP(00) }$  candidates within the window
\unit[$5.11<\mmiss<5.31$]{\gevcc} and
\unit[$-150<\mb - m_B^{PDG}<150$]{\mevcc},
where $m_B^{PDG}$ represents the nominal \Bz~ mass, 
reported by the Particle Data Group~\cite{Hagiwara:fs}.

The sample of \Bztoksksks{} candidates is dominated by random
$\KS\KS\KS$ combinations from $\epem\to\qqbar$ $(\q=u,d,s,c)$
fragmentation. Monte Carlo (MC) studies show that contributions from
other \B{} meson decays are small. 
We exploit topological observables to discriminate the jet-like $\epem\to\qqbar$ events
from the more uniformly distributed \BB{} events. In the $\Y4S$
rest frame we compute the angle $\theta^*_T$ between the thrust
axis of the     $B_{\CP(+-)}$   ($B_{\CP(00)}$) candidate and that of the remaining particles in
the event.  While $|\cos\theta^*_T|$ is highly peaked near 1 for
$\epem\to\qqbar$ events, it is nearly uniformly distributed for
\BB{} events.  We require $|\cos\theta^*_T|<0.9$~$(0.95)$, reducing by one 
order of magnitude the number of background events.
The maximum-likelihood fit described below
also uses discriminant variables 
based on the momenta and angles of tracks in the event to discriminate  $B_{ \CP } $   candidates from $\qqbar$.
They are combined in a Fisher discriminant  $({\cal F})$~\cite{ref:k0spi0prl} for $B_{ \CP (+-) } $ candidates, while
in the case of $B_{\CP(00)}$ candidates we calculate the ratio $L_{2}/L_{0}$ of two Legendre monomials, 
defined as $L_j\equiv\sum_i |{\bf p}^*_i| |\cos \theta^*_i|^j$, where ${\bf p}^*_i$ is the momentum of particle $i$ in the \epem{} rest frame,
$\theta^*_i$ is the angle between ${\bf p}^*_i$ and the thrust axis of
the \B{} candidate and the sum runs over all reconstructed particles
except for the \B{}-candidate daughters.

After all selection requirements are applied, the average $B_{\CP}$ candidate multiplicity in events
with at least one $B_{\CP(00)}$ candidate is approximately $1.67$, coming from 
multiple $\KS \to \ppz$ combinations. In these events, we select the candidate
with the smallest $\chi^2=\sum_{i}
(m_i-m_{\KS})^2/\sigma^2_{m_i}$, where $m_i$ ($m_{\KS}$) is the measured
(nominal $\KS$) mass and $\sigma_{m_i}$ is the estimated uncertainty on
the mass of the $i$th $\KS$ candidate. In simulated events, this selection criterion gives the right
answer about $81\%$ of the time. The remaining misreconstructed events,
coming from fake $\KS \to \ppz$ candidates, do not affect the determination
of $\Delta t$ and have a small impact on the other variables used in the final fit.
The largest correlation is $\sim 2.5\%$. 
 In the case of  events with  $B_{\CP(+-)}$ candidates,  only \unit[$1.4$]{\%} of them  have more than one candidate, 
and we apply the same criterion to select the best combination.

Events coming from $b\to c \bar c s$ would reduce any sensitivity to
departures from the Standard Model as this process is characterized by a Standard Model 
\CP\ asymmetry ($S \sim \stwob$ and $C \sim0$).  We therefore remove all    $B_{ \CP(+-) } $ ($B_{ \CP(00) } $) 
candidates that have a $\KS\KS$ mass combination within $3 \sigma$ ($2 \sigma$)
 of the $\chi_{c0}$ or $\chi_{c2}$ mass. 
After these vetoes the average efficiency, including \KS~ sub branching fractions,
is about 6\% for $B_{ \CP(+-) }$ candidates and 
about 3\%  for $B_{ \CP(00) }$.

Combinatorics from other $B \bar B$ decays constitute a further source of 
background for $B_{ \CP(00) }$ events.
We take this into account by adding a component in the 
likelihood fit (see Sec.~\ref{sec:MLfit}), where the shape of each 
likelihood variable is determined from a simulation of inclusive $B$ decays.
This contribution is found to be negligible in the case of $B_{ \CP(+-) }$ events,
and such a component is not included in the maximum likelihood fit.

\section{Flavor tagging and \deltat reconstruction}
\label{sec:flavdeltat}

For each $B_{\CP}$  candidate we examine the remaining tracks
in the event to determine the decay vertex position
and the flavor of the \Btag candidate. 
 
We use a neural network to determine the flavor of the $\Btag$
meson from kinematic and particle-identification
information~\cite{ref:sin2betaPRL02}. 
Each event is assigned to one of
six mutually exclusive tagging categories, designed to combine flavor
tags with similar performance and \deltat\ resolution.  We
parameterize the performance of this algorithm with a data sample
($B_{\rm flav}$) of fully reconstructed $\Bz\to D^{(*)-}
\pip/\rho^+/a_1^+$ decays. The effective tagging efficiency
obtained from this sample is $Q\equiv\sum_c \epsilon^c
(1-2w^c)^2=0.305\pm 0.004$, where $\epsilon^c$ and $w^c$ are the
efficiencies and mistag probabilities, respectively, for events tagged
in category $c$. 

We compute the proper-time difference $\deltat=(\zrec-\ztag)/\gamma\beta c$
using the known
boost of the \epem{} system and the measured
$\deltaz=\zrec-\ztag$, the difference of the reconstructed decay
vertex positions of the  $B_{\CP}$  and \Btag{} candidates along
the boost direction ($z$).  
A description of the inclusive reconstruction of the \Btag{} vertex 
is given in Ref.~\cite{ref:Sin2betaPRD}.  

To reconstruct the  $B_{\CP}$ vertex from
the \KS{} trajectories we exploit the knowledge of the average
interaction point (IP), which is determined from
the spatial distribution of vertices from two-track events.  We
compute \deltat{} and its uncertainty from a geometric fit to the
$\Ups(4S)\to\Bz\Bzb$ system that takes this IP constraint into
account. We further improve the sensitivity to \deltat{} by
imposing a Gaussian constraint on the sum of the two $B$ decay times
($t_{\CP}+t_{\mbox{tag}}$) to be equal to $2\:\tau_{\Bz}$ with an
uncertainty of $\sqrt{2}\; \tau_{\Bz}$, 
where $\tau_{\Bz}$ is the world average on the $\Bz$ mean life~\cite{Hagiwara:fs},
which effectively constrains the
two vertices to be near the \Y4S{} line of flight~\cite{ref:k0spi0prl}. 
 The uncertainty on the IP position, which
follows from the size of the interaction region, is on the order of
\unit[100]{$\mu$m} horizontally and roughly \unit[4]{$\mu$m} vertically.  
The mean uncertainty on \zrec, a convolution of the interaction 
region and the vertex of the   $B_{\CP}$ decay, is 75{\microns}.
The mean uncertainty on \ztag\ is about 200{\microns} and thus the 
uncertainty in $\deltaz$ is dominated by the uncertainty in the 
vertex of the tagging decay. The resulting resolution
is comparable to that in \Bz\to\jpsi\KS~\cite{ref:k0spi0prl}.

Simulation studies show that the procedure we use to determine the vertex for a
  $B_{\CP}$ decay
provides an unbiased estimate of \zrec{}. The estimate of the
$\deltat$ error in an event reflects the strong dependence of the \zrec{}
resolution on the number of SVT
layers traversed by the \KS\ decay daughters. However, essentially all
events have at least one \KS\ candidate for which both tracks have at least
one hit in the inner three SVT layers (at radii from
\unit[$3.2$]{cm} to \unit[$5.4$]{cm}). In this case the mean
\deltat\ resolution is comparable to that in decays in which the
vertex is directly reconstructed from charged particles
originating at the $B$ decay point~\cite{ref:Sin2betaPRD}.  For a
small fraction (0.1\%) of the signal events, at least one \KS\ has tracks
with hits in the outer two SVT layers (at radii \unit[$9.1$]{cm}
to \unit[$14.4$]{cm}) but none of the three \KS\ s have hits in the inner three layers. In this case 
the resolution is nearly two times worse but the event can still be used in the \CP fit. 
For events with \unit[$\sigma_{\deltat}>2.5$]{ps} or \unit[$|\deltat|>20$]{ps},
the \deltat{} information is not used. However, since \cksksks~ can also be extracted from flavor tagging
information alone, these events still contribute to the measurement of \cksksks.

The \deltat~ resolution function ${\cal R}$ is parameterized as the sum of a `core' and a
`tail' Gaussian distribution, each with a width and mean proportional to 
$\sigma_{\deltat}$, and a third Gaussian with a mean of
zero and a width fixed at \unit[$8$]{ps}~\cite{ref:Sin2betaPRD}.  
We have verified on data that the parameters of ${\cal R}$ for  $B_{\CP}$ decays are similar to 
those obtained from the $B_{\rm flav}$ sample, even when the IP constrained vertexing technique is applied.
Therefore, we extract these parameters from a fit to the $B_{\rm flav}$ sample.  
We find that the \deltat{} distribution of background candidates is well described by a
delta function convolved with a resolution function having the same
functional form as that for the signal. The parameters of the
background function are determined in the fit.

\section{Maximum Likelihood fit }
\label{sec:MLfit}

We extract the results from unbinned maximum-likelihood 
fits to the kinematic, event shape, and $\deltat$ variables.
For each subsample we consider  the logarithm of an extended likelihood function
\begin{eqnarray}
{{\cal L} = 
e^{-(\sum_j^n N_j)}
\times}  
{\prod_i^{N_T}{
}
   \sum_j^n  N_j 
{\cal P}^i_{j}  ,}  
\nonumber
\end{eqnarray}
\noindent where 
${\cal P}_j$ is  the probability density function (PDF) for the j$^{th}$ fit 
component and $N_j$  the event yields of each component   
($N_S$ signal events, $N_{\qqbar}$   $\qqbar$ events  and, for $B _ { \CP(00) } $ only,  $N_{B \bar B}$ 
$B \bar B$ decay events)
$N_T$ is the total number of events  selected.
The two   ${\cal L}$  are then summed and maximized to determine the common \sksksks~ and \cksksks~ \CP asymmetry 
parameters  and the $N_j$ which are specific to each subsample.

The  \deltat  PDF for a given tagging category is 
${\cal P}^c(\deltat,\sigma_{\deltat})\epsilon^c$
where $\epsilon^c$ is the tagging efficiency for tag category $\c$.  
The total likelihood $\cal L$ is the product of
likelihoods for each tagging category, 
and the free parameters are determined by maximizing the quantity $\ln \cal L$.  
Along with the \CP\ asymmetries \sksksks~ and \cksksks, 
the fit extracts $\epsilon^c$ and other parameters for background.  
The background PDFs include parameters for the \deltat-resolution function ${\cal R}$ 
and for asymmetries in the rate of \Bz{} versus \Bzb{} tags.  
We extract  43  parameters from the fit.

The observables are sufficiently uncorrelated
that we can construct the likelihoods as the products of one-dimensional PDFs.
The signal PDFs are parameterized from signal MC events.  
For background PDFs we
determine the functional form from data in the sideband regions of the
other observables where backgrounds dominate.  
We include these regions in the fitted sample 
and simultaneously extract the parameters
of the background PDFs along with the fit results.

\begin{figure}[!tbp]
\begin{center}
\includegraphics[width=0.49\linewidth]{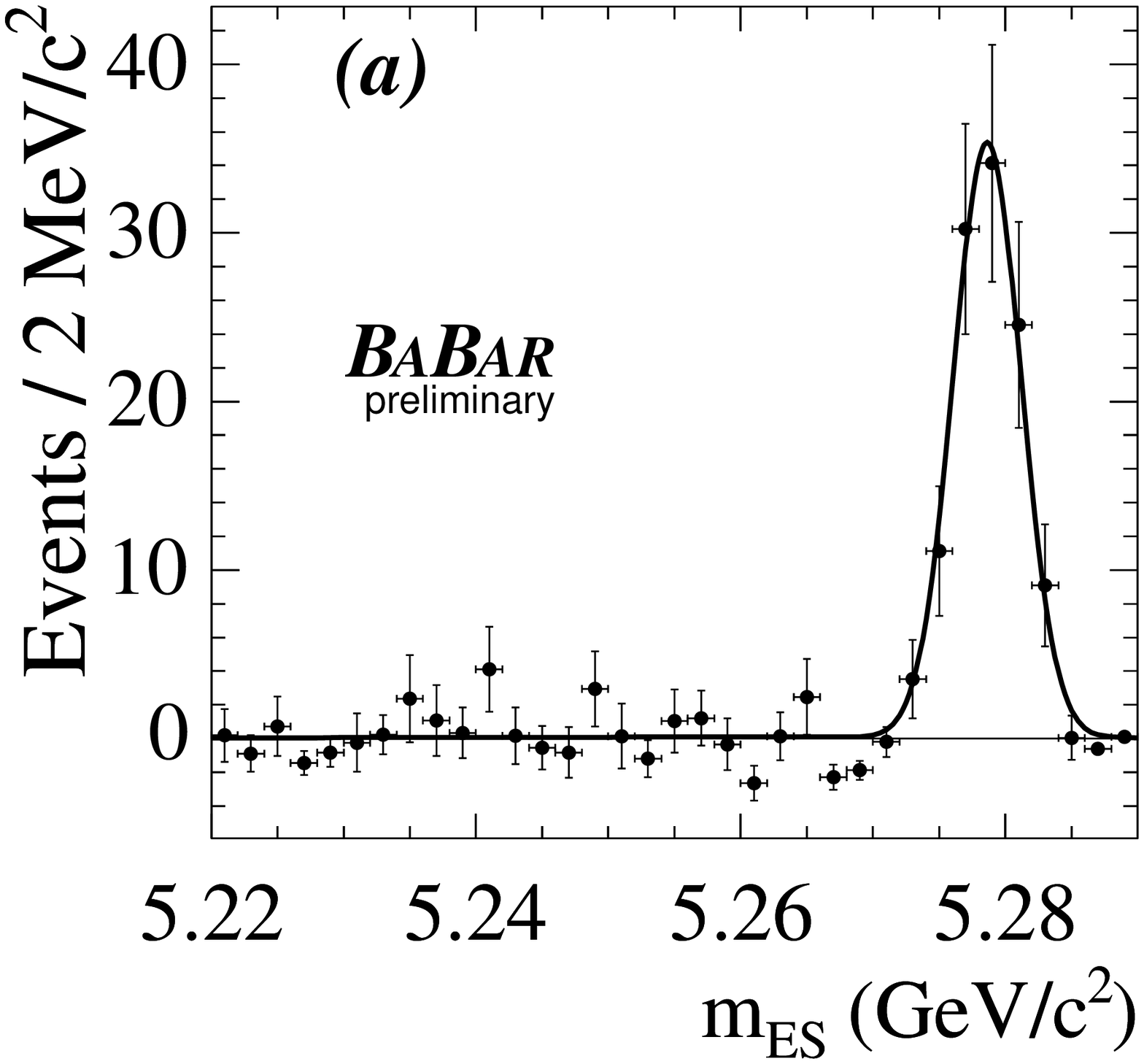}
\includegraphics[width=0.49\linewidth]{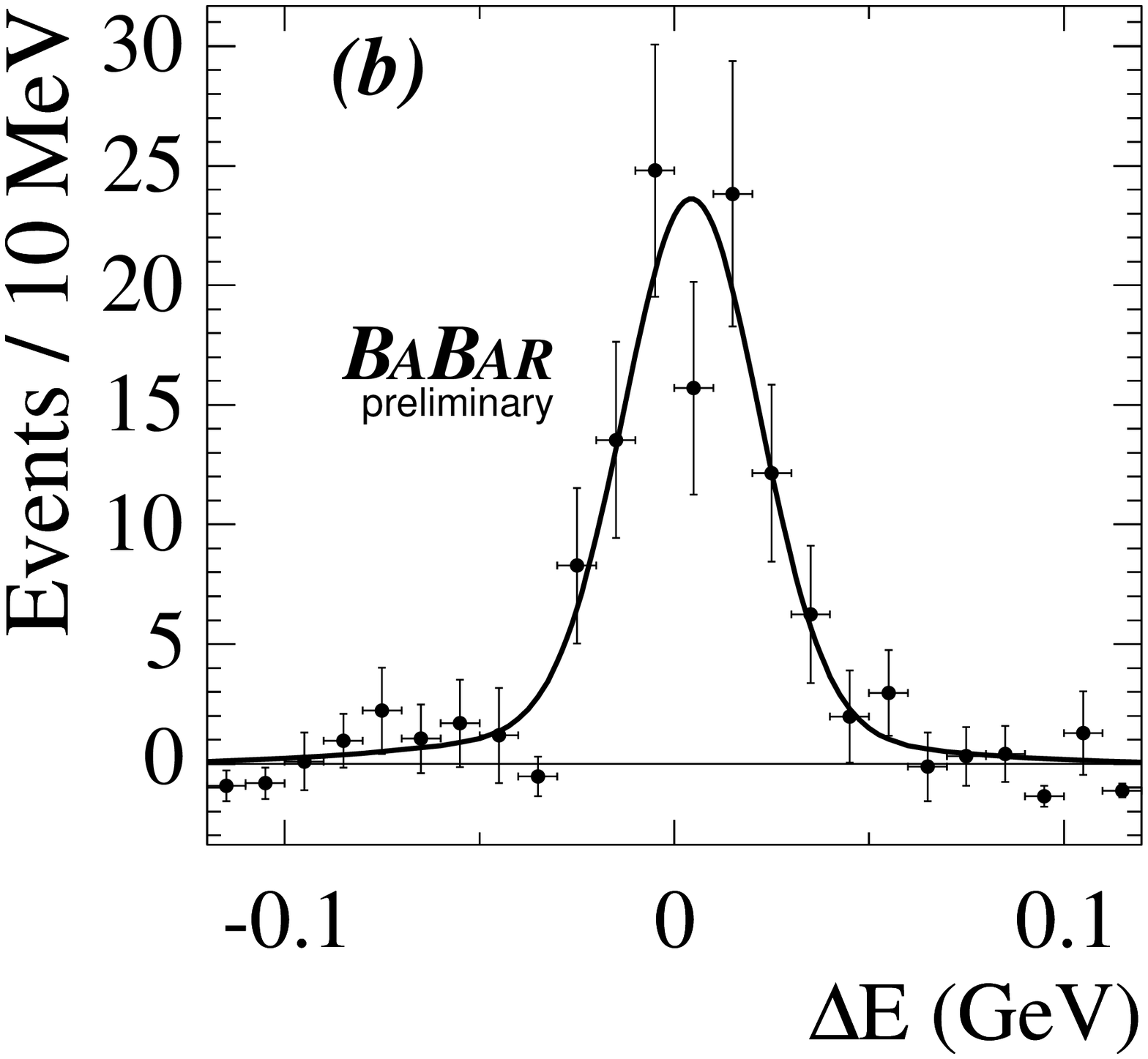} 
\includegraphics[width=0.49\linewidth]{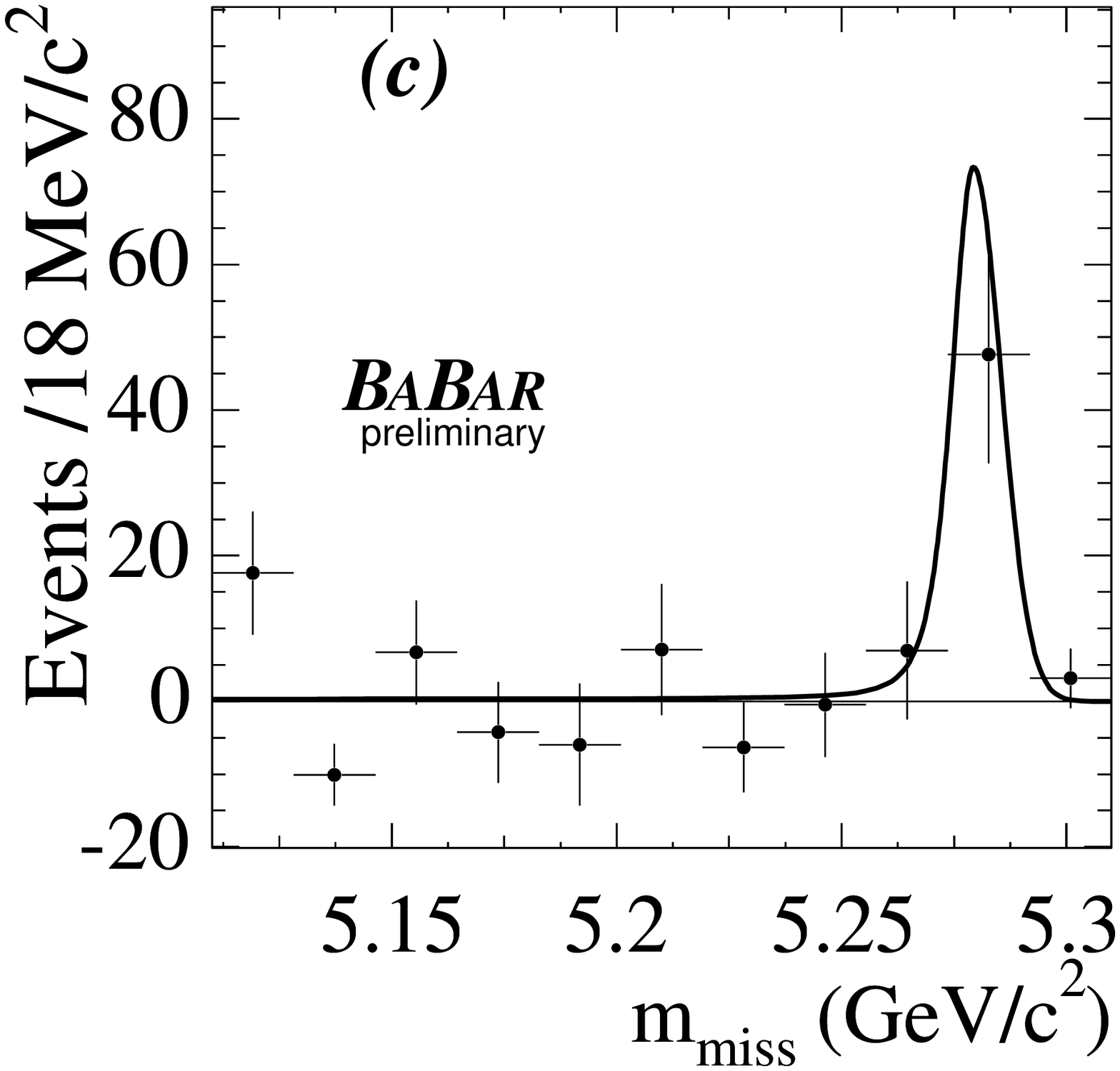}
\includegraphics[width=0.49\linewidth]{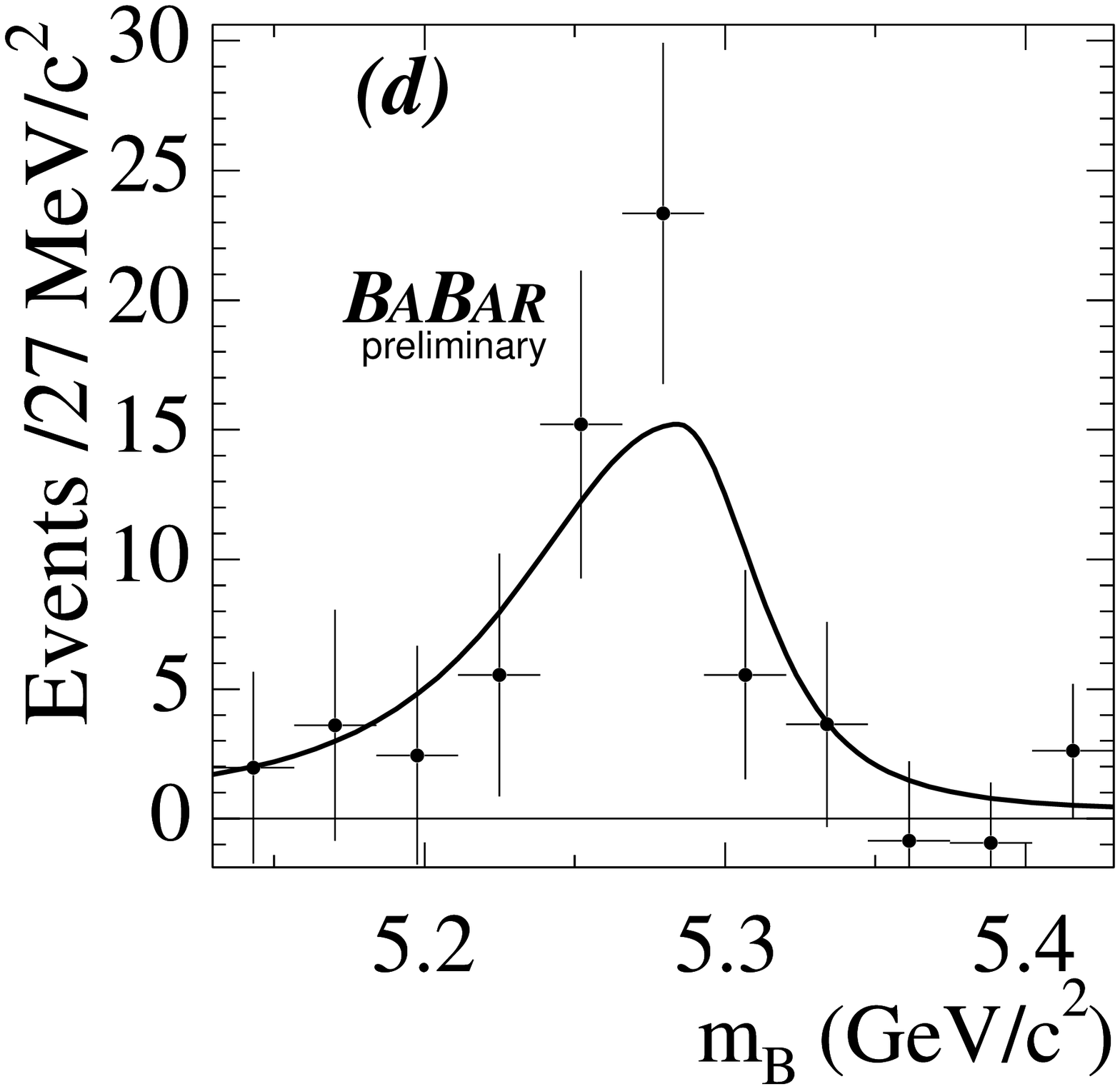}
\caption{sPlots of (a) $\mes$ and (b) $\DeltaE$ for  $B_{\CP(+-)}$ subsample
 and of (c) $\mmiss$ and (d) \mb \ for $B_{\CP(00)}$ subsample.} 
\label{fig:spplots}
\end{center}
\end{figure}

There are  786 $B_{\CP(+-)}$   and   4550   $    B_{\CP(00)}$ candidates that pass all the selection criteria. 
In Table~\ref{tab:summres} the events yields obtained in the fit are summarized for the two subsamples  separately. 
Figure~\ref{fig:spplots} shows the \mes{} and $\DeltaE$  ($\mmiss$ and  \mb ) distributions for $B_{\CP(+-) }$ ($B_{\CP(00) }$) signal events  
with the sPlot event weighting technique~\cite{splot}.  The results  of the fit  are  plotted as curves.

\section{Results}
\label{sec:Physics}
 The \CP\ parameters extracted from the fit are summarized in Table 1. We obtain
 \begin{eqnarray*}
\sksksks  &=&  -0.66 \pm 0.26 \pm 0.08,  \\
 \cksksks &=& -0.14 \pm 0.22  \pm 0.05, 
 \end{eqnarray*}
 where the first error is statistical and the second systematic (evaluated as described 
 in the following Section).  
The correlation between \sksksks~ and \cksksks~ is -8.5\%.
We evaluate the statistical  significance of \CP violation to be $2.6\sigma$ 
 by calculating the $2 \Delta log {\cal L}$ variation when fitting  
data with  \sksksks~ and \cksksks~ fixed to zero. 
 Using a Monte Carlo technique, in which we assume that the measured values for the \CP~ parameters
 on the combined data sample are the true values, we evaluate the probability of measuring the values 
 reported in Table~\ref{tab:summres} for the two sub-samples.
 We find that the two sub-samples agree  within  $1.6 \sigma$.

\begin{figure}[!tbp]
\begin{center}
\parbox{0.47\textwidth}{\includegraphics[width=1.2\linewidth]{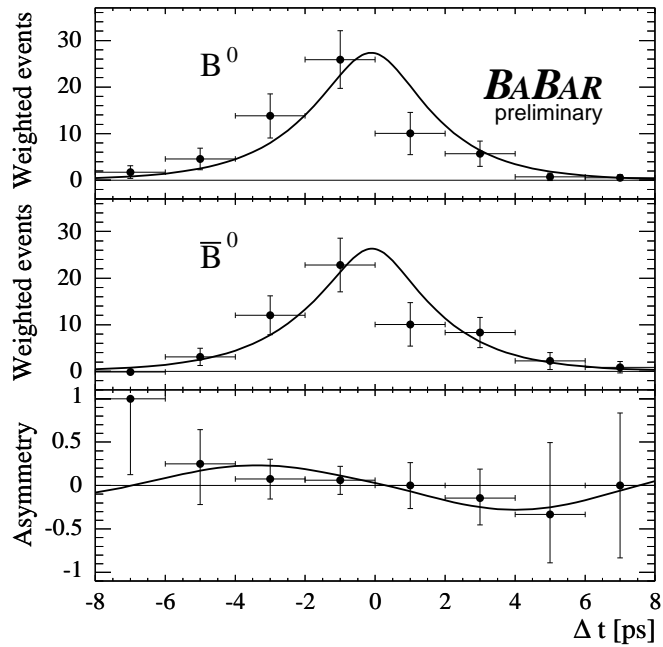}}
\end{center}
\caption{
  Distributions of $\deltat$ for events weighted with the sPlot technique for
  $B_{\rm tag}$ tagged as  $\Bz$  (top) or   $\Bzb$ (center), and  the
  asymmetry ${\cal A}(\deltat)$ (bottom).  The points are weighted data
  and the curves are the corresponding PDF projections. }
\label{fig:dtplot}
\end{figure}

Figure~\ref{fig:dtplot} shows distributions of
$\deltat$ for $\Bz$-tagged and $\Bzb$-tagged events, and the asymmetry
${\cal
  A}(\deltat) = \left( N_{\Bz} -
  N_{\Bzb}\right) /\left( N_{\Bz} + N_{\Bzb}\right) ,$ 
obtained with the sPlot event weighting technique~\cite{splot}.
This result is in good agreement with the value of  $\stwob$  
from $b \to c\cbar\s$ decays~\cite{unknown:2006bi}.

\begin{table}[htb]
   \centerline{\small
    \begin{tabular}{lcccc}
      \hline\hline
      &     $B_{\CP(+-)}$   & $ B_{\CP(00)}$   & Combined  \\ 
      \hline
      $N_S$                   & 116 $\pm$ 12             &   60 $\pm$ 12      & $-$ \\ 
      $N_{\qqbar}$      &     670 $\pm$ 26           &   4482 $\pm$ 71   & $-$ \\ 
      $N_{B \bar B}$   &    $-$      &   8 $\pm$ 25   & $-$ \\ 
      \hline
      \sksksks                     &  -1.04 $^{+0.26}_{-0.17}$  &    0.37 $\pm$ $^{+0.52}_{-0.54}$   &  -0.66 $\pm$ 0.26 \\
      \cksksks                     &  -0.31 $^{+0.25}_{-0.23}$    &   0.21 $\pm$ 0.38  &  -0.14 $\pm$ 0.22 \\ 
      \hline\hline
    \end{tabular}}
  \caption{Events yields and \CP\ asymmetry parameters obtained in the fit. Statistical errors  only are shown.}
  \label{tab:summres}
\end{table}

\section{Systematic studies}
\label{sec:Systematics}

We obtain systematic uncertainties in the \CP\ coefficients \sksksks~ and \cksksks~ due to
the parameterization of kinematic variables and event shape  PDFs in signal and background by
varying the parameters within one standard deviation (evaluated from a fit to Monte Carlo simulated events). 
There might be a contribution to \cksksks~
and \sksksks~ from \CP violation in the $B \bar B$ background.
In the fit, the values of the effective \CP parameters ($S_{B \bar B}$ and $C_{B\bar B}$) for the $B\bar B$ 
background are fixed to zero.
They are varied within the whole physical allowed range $S^2_{B\bar B}+C^2_{B\bar B}\leq 1$
and we take the largest variation on signal \sksksks~ and \cksksks~ as systematic uncertainty.

We evaluate the uncertainties associated with
the assumed parameterization of the \deltat\  resolution function for signal
and \BB-background by varying the parameters within one standard deviation 
(extracted from a fit to the \Bflav\ sample).  
The uncertainties due to knowledge of  efficiencies and dilutions of  flavor tagging  
and possible difference in the efficiency between \Bz\ and \Bzb are  evaluated in the same way. 
 The  mass difference between the two \Bz\ mass eigenstates, \deltamd, 
and the \Bz~ mean life, $\tau_{B^0}$, values held fixed in the fit are varied within their uncertainties 
determined in world averages \cite{Hagiwara:fs}.

We also estimate different uncertainties associated 
with vertexing. We take the largest value of $\sksksks(\cksksks)_{\rm{fit}}-\sksksks(\cksksks)_{\rm{true}}$ from fits 
to signal Monte Carlo events where
realistic misalignments of the SVT silicon wafers have been introduced.
Here the $\sksksks(\cksksks)_{\rm{fit}}$ represents the result of the fit to these simulated events,
while $\sksksks(\cksksks)_{\rm{true}}$ represents input values in the 
Monte Carlo generation. 
We include an additional contribution from the comparison of
the description of the resolution function (RF)  between  
IP-constrained vertexing and nominal vertexing in the case of 
$B^0 \to J/\psi K^0_S$ events.

We assign a systematic uncertainty on our knowledge of 
the beam spot position by shifting the beam position in the simulation
by $\pm 20~\mu m$ in the vertical direction. 
The sensitivity due to any calibration problems or time-dependent effects is evaluated by 
smearing the beam-spot position by an
additional $\pm20~\mu m$ in the vertical direction.
The effect of neglecting possible correlations between the variables
in the fit is estimated with a Monte Carlo technique.
We also estimate  the errors due to the effect of doubly
CKM-suppressed decays on the tag side by varying the value of the rate of such decays and the  
strong and weak  phase within conservative limits~\cite{ref:tagint}.

We add these contributions in quadrature to obtain the total
systematic uncertainty. The summary is reported in Table 2.
The largest contributions are related to the knowledge of the PDF parameters
and the \CP content of the $B\bar B$ background.

\begin{table}[hbtp]
\begin{center}
\begin{small}
\begin{tabular}{ccccc} \\ \hline\hline
      & $\Delta~S(+)$ & $\Delta~S(-)$ & $\Delta~C(+)$ & $\Delta~C (-)$  \\
\hline

$(+-)$ pdf parameters           & 0.007 & 0.010 & 0.020 & 0.018 \\
$(00)$ pdf parameters          & 0.025 & 0.022 & 0.027 & 0.022 \\
$B\bar B$ \CP                         & 0.077 &0.077  &0.026  & 0.026\\
data  RF                                  & 0.004 & 0.006 & 0.008 & 0.006 \\
flavor tagging                       & 0.007 & 0.010 & 0.017 & 0.012 \\
$\tau_{B}$ and \deltamd\        & 0.004 & 0.003 & 0.005 & 0.009 \\
SVT alignment                    & 0.016 & 0.016 & 0.008 & 0.008 \\
vertexing method               & 0.016 & 0.016 & 0.003 & 0.003 \\
beam-spot               & 0.004 & 0.004 & 0.001 & 0.001 \\ 
fit correlations                           &  0.004  &  0.004  &  0.025  & 0.025  \\
tag side interference   & 0.001 & 0.001 & 0.011 & 0.011 \\\hline
total errors                     &0.085   &0.085  &0.055 &0.051 \\
\hline\hline
\end{tabular}
\end{small}
\caption{Summary of systematic uncertainties on S and C.\label{tab:systsummary}}
\end{center}
\end{table}

\section{Conclusions}
\label{sec:Summary}

In summary, we have measured  the time-dependent \CP-violating  asymmetries 
for the decay  \Bztoksksks:   \mbox{$\sksksks = -0.66\pm 0.26 \pm 0.08$} and 
\mbox{$\cksksks = -0.14 \pm 0.22 \pm 0.05$}.    
Within the current experimental uncertainties, these  measurements are in good agreement with the SM expectation.

\section{Acknowledgements}
\label{sec:Acknowledgments}
We are grateful for the 
extraordinary contributions of our \pep2\ colleagues in
achieving the excellent luminosity and machine conditions
that have made this work possible.
The success of this project also relies critically on the 
expertise and dedication of the computing organizations that 
support \babar.
The collaborating institutions wish to thank 
SLAC for its support and the kind hospitality extended to them. 
This work is supported by the
US Department of Energy
and National Science Foundation, the
Natural Sciences and Engineering Research Council (Canada),
Institute of High Energy Physics (China), the
Commissariat \`a l'Energie Atomique and
Institut National de Physique Nucl\'eaire et de Physique des Particules
(France), the
Bundesministerium f\"ur Bildung und Forschung and
Deutsche Forschungsgemeinschaft
(Germany), the
Istituto Nazionale di Fisica Nucleare (Italy),
the Foundation for Fundamental Research on Matter (The Netherlands),
the Research Council of Norway, the
Ministry of Science and Technology of the Russian Federation, 
Ministerio de Educaci\'on y Ciencia (Spain), and the
Particle Physics and Astronomy Research Council (United Kingdom). 
Individuals have received support from 
the Marie-Curie IEF program (European Union) and
the A. P. Sloan Foundation.


\begin{thebibliography}{99}
\bibitem{ref:cc} Unless explicitly stated otherwise, conjugate decay modes are
assumed throughout this paper.
\end{thebibliography}
\end{document}